\documentclass[conference]{IEEEtran}
\usepackage{graphicx}
\usepackage{multirow} 
\begin{document}
%
\title{Simulation technique for available bandwidth estimation}

\author{\IEEEauthorblockN{T.G. Sultanov}
\IEEEauthorblockA{Samara State Aerospace University\\
Moskovskoe sh., 34,\\
Samara, 443086, Russia\\
E-mail: tursul@rambler.ru}
\and
\IEEEauthorblockN{A.M. Sukhov}
\IEEEauthorblockA{Samara State Aerospace University\\
Moskovskoe sh., 34,\\
Samara, 443086, Russia\\
E-mail: amskh@yandex.ru}}

\maketitle

\begin{abstract}
The paper proposes a method for measuring available bandwidth, based on testing network packets of various sizes (Variable Packet Size method, VPS). The boundaries of applicability of the model have been found, which are based on the accuracy of measurements of packet delays, also we have derived a formula of measuring the upper limit of bandwidth. The computer simulation has been performed and relationship between the measurement error of available bandwidth and the number of measurements has been found. Experimental verification with the use of RIPE Test Box measuring system has shown that the suggested method has advantages over existing measurement techniques. \textit{Pathload} utility has been chosen as an alternative technique of measurement, and to ensure reliable results statistics by SNMP agent has been withdrawn directly from the router.
\\
\end{abstract}


\IEEEpeerreviewmaketitle

\section{Introduction}
Various real-time applications in the Internet, especially transmission audio and video information, become more and more popular. The major factors defining quality of the service are quality of the equipment (the codec and a video server) and  available bandwidth in Internet link. ISP providers should provide the required bandwidth for voice and video applications to guarantee the submission of demanded services in the global network.

In this paper, network path is defined as a sequence of links (hops), which forward packets from the sender to the receiver. There are various definitions for the throughput metrics, but we will use the approaches accepted in \cite{s5,s9,s10}.

\begin{figure}
\centering
\includegraphics[height=2.525cm]{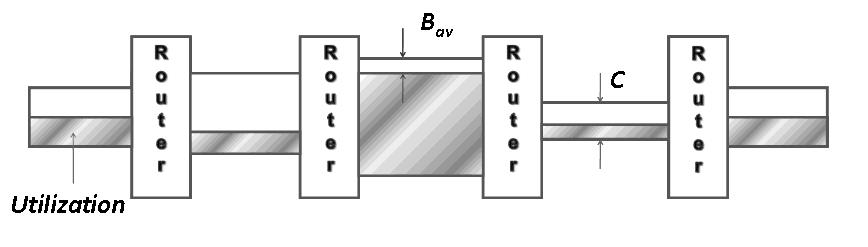}
\caption{Illustration of throughput metrics}
\label{f1}
\end{figure}

Two bandwidth metrics that are commonly associated with a path are the capacity $C$ and the available bandwidth $B_{av}$ (see Fig~\ref{f1}). The {\em capacity C} is the maximum IP-layer throughput that the path can provide to a flow, when there is no competing traffic load (cross traffic). The {\em available bandwidth} $B_{av}$, on the other hand, is the maximum IP-layer throughput that the path can provide to a flow, given the path's current cross traffic load. The link with the minimum transmission rate determines the capacity of the path, while the link with the minimum unused capacity limits the available bandwidth. Moreover measuring available bandwidth is important to provide information to network applications on how to control their incoming and outgoing traffic and fairly share the network bandwidth.

Another related throughput metric is the {\em Bulk-Transfer-Capacity} (BTC). BTC of a path in a certain time period is the throughput of a bulk TCP transfer, when the transfer is only limited by the network resources and not by limitations at the end-systems. The intuitive definition of BTC is the expected long-term average data rate (bits per second) of a single ideal TCP implementation over the path in question.

In order to construct a perfect picture of a global network (monitoring and bottlenecks troubleshooting) and develop the standards describing new applications, modern measuring infrastructure should be installed. In this paper we are describing usage of RIPE Test Box measurement system, which is widely used \cite{s7}.

According to \cite{s7} this system doesn't measure the available bandwidth, but it collects the numerical values, which characterize key network parameters such as packet delay, jitter, routing path, etc. 

In this paper we attempt to provide universal and simple model that allow us to estimate available bandwidth based on received data from RIPE Test Box measurement infrastructure. The method is based on {\em Variable Packet Size} (VPS) method and was used in \cite{s6}. This method allows us to estimate network capacity of a hop $i$ by using connection between the Round-Trip Time (RTT) and packet size $W$.

\section{The model and its applicability}
The well-known expression for throughput metric describing the relation between a network delay and the packet size is Little's Law \cite{s13}:
 	\begin{equation}
  B_{av}=W/D,
  \label{e1}
	\end{equation}
where $B_{av}$ is available bandwidth, $W$ is the size of transmitted packet and $D$ is network packet delay (One Way Delay). This formula is ideal for calculating the bandwidth between two points on the network that are connected without any routing devices. In general case delay value is caused by constant network factors as propagation delay, transmission delay, per-packet router processing time, etc \cite{s9}.

According to \cite{s1}, Little's Law could be modified with $D^{fixed}$:
 	\begin{equation}
  B_{av}=W/(D-D^{fixed}),
  \label{e2}
	\end{equation}
where $D^{fixed}$ is {\em minimum fixed delay} for the packet size $W$. The difference between the delays $D$ and $D^{fixed}$ is the {\em variable delay component} $d_{var}$. In paper \cite{s3} it was shown that variable delay is exponentially distributed.

Choi \cite{s2}, Hohn \cite{s12} showed that minimum fixed delay component $D^{fixed}(W)$ for the packet size $W$ is an linear (or affine) function of its size:
 	\begin{equation}
  D^{fixed}(W)=W\sum_{i=1}^h 1/C_i + \sum_{i=1}^h \delta_i,
  \label{e3}
	\end{equation}
where $C_i$ is capacity of appropriate link and $\delta_i$ is propagation delay. To prove this assumption authors experimentally found the minimum fixed delays for packets of the same size for three different routes and constructed function of dependence of a delay from the packet size $W$.

In order to eliminate minimum fixed delay $D^{fixed}(W)$ from Eqn.~(\ref{e2}) we are suggesting to test network link with packets of different sizes \cite{s1}, so that the packet size varied at the maximum possible value without possible router fragmentation. Then Eqn.~(\ref{e2}) could be modified to a suitable form for measuring available bandwidth:
 	\begin{equation}
  B_{av}=\frac{W_2-W_1}{D_2-D_1}
  \label{e4}
	\end{equation}

This method allows us to find a way to eliminate the measurement limitations of variable delay component $d_{var}$. The variable delay component is the cause rather large measurement errors of other methods, which will be described in the last section of this paper.

Proposed model is quite simple, but it's still difficult to find accurate measuring infrastructure. The first problem concerns the applicability of the model, i.e., what range of throughput metrics can be measured using with this method. The second issue is number of measurements (group of packets) needed to achieve a given accuracy.

First problem could be solved by using measurement error (based on delay accuracy measurement):
 	\begin{equation}
  \eta=\frac{\Delta B}{B}=\frac{2\Delta D}{D_2-D_1},
  \label{e5}
	\end{equation}
where $\eta$ is relative error of measurement available bandwidth, $\Delta B$ is absolute error of measurement available bandwidth and $\Delta D$ is precision of measuring the packet delay. 

With this expression we can easily find an upper bound \ensuremath{\bar{B}} for available bandwidth:
 	\begin{equation}
  \bar{B}=\frac{W_2-W_1}{2\Delta D}\eta
  \label{e6}
	\end{equation}

Thus, with the RIPE Test Box, which allows to find the delay to within 2 microseconds $\Delta D=2\cdot10^{-6}$ second precision, we can measure the available bandwidth to the upper bound $\ensuremath{\bar{B}}=300$ {\em Mbps} with relative error $\eta=10\%$. Moreover, if we were using standard utility \textit{ping}, with a relative error $\eta=25\%$ and precision of 1 millisecond $\Delta D=10^{-3}$ second, we could get results for network available bandwidth up to $1.5$ {\em Mbps}.

\section{Experimental comparison of different methods}
In this part of the paper we would like to show results based on comparing different methods of measuring available bandwidth by using obtained results of experiments.

The experiment was divided in three stages. In the first stage we have used RIPE Test Box measurement system with two different packet sizes. Number of measurement systems in global measurement infrastructure reaches 80, these points are covering major Internet world's centers of the Internet, reaching highest density in Europe. The measurement error of packet delay is 2-12 $\mu s$ \cite{s14}. In order to prepare the experiments, three Test Boxes have been installed in Moscow, Samara and Rostov on Don in Russia during 2006-2008 years in support of RFBR grant 06-07-89074. For further analysis we collected several data sets containing up to 3000 data results in different directions, including Samara - Amsterdam (tt01.ripe.net - tt143.ripe.net). Based on these data, we calculated available bandwidth and dependence of measurement error on the number of measurements (see Fig~\ref{f3}). 

The second stage was comparing data obtained with our method, with the results of traditional methods of throughputs measurement. \textit{Pathload} was selected as a tool that implements a traditional method of measurement product \cite{s10}. This software is considered one of the best tools to assess the available bandwidth. \textit{Pathload} uses Self-Loading Periodic Streams (SLoPS). It is based on client-server architecture, which is its disadvantage, since you want to install the utility on both hosts. \textit{Pathload} advantage is that it does not require root privileges, as the utility sends only UDP packets. 

The results of measurements \textit{pathload} displayed as a range of values rather than as a single value. Mid-range corresponds to the average throughput, and the band appreciates the change in available bandwidth during the measurements. 

The third stage involves the comparison of data obtained in the first and second stages with the data directly from the router SSAU which serves the narrowest part of the network.

\begin{figure*}
\centering
\includegraphics[height=6cm]{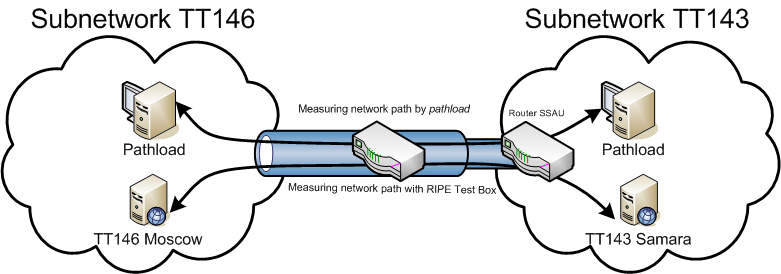}
\caption{Scheme of the experiment}
\label{f2}
\end{figure*}

The experiment between points tt143.ripe.net (Samara State Aerospace University) and tt146.ripe.net (FREENet, Moscow) consists of three parts:
\begin{enumerate}
	\item 	
	Measuring the available bandwidth by testing pairs of packets of different sizes using the measuring system RIPE Test Box (packet size of 100 and 1100 bytes);
	\item 
	Measurements of available bandwidth using the utility \textit{pathload};
	\item 
	Measuring the available bandwidth by MRTG on the router SSAU which serves the narrowest hop of routing (see Fig~\ref{f2}).
\end{enumerate}

It is worth noting that all the three inspections should be conducted simultaneously in order to maximize the reliability of the statistics. The structure of the measuring system RIPE Test Box meets all the requirements of our method - it allows to change the size of the probe packet and find high-precision delay.

By default, the test packet size is 100 bytes. There are special settings that allow adding testing packets of up to 1500 bytes to the desired frequency. In our case it is reasonable to add a packet size of 1100 bytes. It should be noted that testing of these packets does not begin until the next day after sending a special request.

In order to gain access to the test results it is necessary to apply for remote access (\textit{telnet}) to the RIPE Test Box on port 9142. The data includes information about the desired delay packets of different sizes. In order to extract the data it is necessary to identify the packet on receiving and transmitting sides.

First, it should to explore sender's side:
\begin{table}[!h]
	\centering
		\begin{tabular}{ll}
SNDP 9 1263374005 -h tt01.ripe.net -p 6000 -n 1024 -s & 1353080538\\
SNDP 9 1263374005 -h tt146.ripe.net -p 6000 -n 100 -s & \textbf{1353080554}\\
SNDP 9 1263374005 -h tt103.ripe.net -p 6000 -n 100 -s & 1353080590\\
		\end{tabular}
	\caption{The data of sending box}
	\label{t1}
\end{table}

The last value in line is the serial number of packet. It should to be remembered to get a packet already on the receiving side of the channel. Below is a sample line on the receiving side.
\begin{table}
	\centering
		\begin{tabular}{ll}
		{\scriptsize RCDP 12 2 89.186.245.200 55730 193.233.1.69 6000} & {\scriptsize 1263374005.779364}\\ 
		{\scriptsize \textbf{0.009001} 0X2107 0X2107 \textbf{1353080554} 0.000001 0.000001}\\
		\\
		{\scriptsize RCDP 12 2 200.19.119.120 57513 193.233.1.69 6000} & {\scriptsize 1263374005.905792}\\ 
		{\scriptsize 0.160090 0X2107 0X2107 1353080554 0.000003 0.000001}\\
		\end{tabular}
	\caption{The data of receiving box}
	\label{t2}
\end{table}

For a given number of packet is easy to find the packet delay. In this case it is 0.009001 sec. The following packet 1353091581 is the size of 1100 bytes and the delay is 0.027033 seconds. Thus, the difference is 0.018032 seconds. Other values are processed packet delay similar. 

The mean value $D_2-D_1$ should be used in Eqn.~(\ref{e4}), so it's necessary to average several values, going consistently. In the present experiment, the averaged difference $D_{av}(1100)-D_{av}(100)$ amounted to 0.000815 seconds in the direction $tt143\rightarrow tt146$. Then the available bandwidth can be calculated as:

\[B_{av}(tt143\rightarrow tt146)=\frac{8\times 1000}{0.000815}=9.8 \textit{ Mbps}\]
The average difference in the direction $tt146\rightarrow tt143$ was 0.001869 seconds. Then the available bandwidth will be:

\[B_{av}(tt146\rightarrow tt143)=\frac{8\times 1000}{0.001869}=4.28 \textit{ Mbps}\]

Measuring of \textit{pathload} utility was with periodically troubles, even though that had been opened all necessary ports. In the direction of measurement $tt146\rightarrow tt143$ the program has not get any results despite all our attempts. It is idle and filled the channel chain packets. The \textit{pathload} results give a large spread of values, clearly beyond the capacity of the investigated channel. The other measurements with utilities \textit{pathChirp} and \textit{IGI} were also unsuccessful. Programs give errors and refused to measure the available bandwidth. 

Therefore, it was decided to compare the results obtained by different methods with data obtained directly from the router. \textit{Traceroute} utility determines "bottleneck" of route path between SSAU and Institute of Organic Chemistry at the Russian Academy of Sciences. It was an external SSAU router which bandwidth was limited up to $30$ {\em Mbps}. SNMP agent collects statistic of the border router SSAU.

All data are presented in Table~\ref{t3} indicating the time of the experiment.
\begin{table*}
	\centering
		\begin{tabular}{|c|c|c|c|c|c|}
		\hline
N	& Date	& Direction	& Available bandwidth	& Available bandwidth	& Data from router
\\ & & & (data of RIPE Test Box) & (data of \textit{pathload}) & \\
		\hline
1	& 13.01.2010	& $tt143\rightarrow tt146$	& $10.0\pm2.2$ \em {Mbps}	& $21.9\pm14.2$ \em {Mbps}	& $12.1\pm2.5$ \em {Mbps}\\	
		\hline	
2	& 13.01.2010	& $tt146\rightarrow tt143$	& $4.4\pm1.2$ \em {Mbps}	&  & $7.8\pm3.8$ \em {Mbps}\\	
		\hline
3	& 23.01.2010	& $tt143\rightarrow tt146$	& $20.3\pm5.1$ \em {Mbps}	& $41.2\pm14.0$ \em {Mbps}	& $18.7\pm1.1$ \em {Mbps}\\	
		\hline	
4	& 23.01.2010	& $tt146\rightarrow tt143$	& $9.3\pm2.7$ \em {Mbps}	&  & $11.3\pm2.6$ \em {Mbps}\\	
		\hline
5	& 06.02.2010	& $tt143\rightarrow tt146$	& $9.2\pm1.4$ \em {Mbps}	& $67\pm14$ \em {Mbps}	& $12.0\pm2.0$ \em {Mbps}\\	
		\hline	
6	& 06.02.2010	& $tt146\rightarrow tt143$	& $3.5\pm1.2$ \em {Mbps}	&  & $4.5\pm2.0$ \em {Mbps}\\	
		\hline
		\end{tabular}
	\caption{Comparative analysis of measurement results}
	\label{t3}
\end{table*}
Table~\ref{t3} shows that the results obtained by our method and router data are in a good agreement, while the \textit{pathload} measurements differ. The study of the statistical type of delay \cite{s3} provides an answer to the question why this is happening. The dispersion of measurements results speaks presence of a variable part of delay $d_{var}$. This utility uses Self-Loading Periodic Streams (SLoPS) like most others. This method consists in the generation of packets chain with redundant frequency when time packet delivery will significantly increase due to long queues at the routers. In this case, the transmitter starts to reduce the frequency of packets generation until the queue it disappears. Next, the process will be repeated for as long as the average frequency of packets generation will not approach the available bandwidth. The main disadvantage of this technique is unreliable measurements because they have not considered the influence of the variable part of delay. This is the reason for fantastic $90$ {\em Mbps} \textit{pathload} result for channel with a $30$ {\em Mbps} capacity.

\section{The required number of measurements}
The main disadvantage of most modern tools is a large spread of values of available bandwidth. Measurement mechanisms of throughput utilities do not take into account the effect of the variable part of the delay.  Unfortunately, in all developed utilities compensations mechanism for the random component's delay isn't provided. 

\begin{figure*}
\centering
\includegraphics[height=8cm]{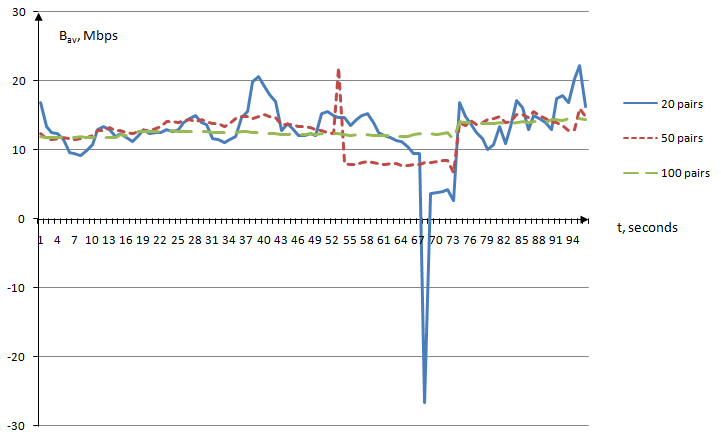}
\caption{Dependence of available bandwidth on the number of measurements}
\label{f3}
\end{figure*}

Any method that gives accurate results should contain mechanism for smoothing the impact $d_{var}$. In order to understand the effect of the variable part on the measurement results we turn to the following experiment. The series of measurements have been made between RIPE Test Boxes: tt01.ripe.net (Amsterdam, Holland) and tt143.ripe.net (Samara State Aerospace University, Russian Federation). It was received about 3000 values of delay packet size of 100 and 1024 bytes in both directions. Using the presented method quantities of available bandwidth have been calculated for cases where the averaging is performed on 20, 50 and 100 pairs of values. On Fig.~\ref{f3} the schedule of the available bandwidth calculated for various conditions of averaging is represented.

Apparently from the schedule, beatings of the calculated available bandwidth remain critical at 20 averaged values. At 50 it is less noticeable, and at 100 the values the curve is almost equalized. There was a clear correlation between the number of measurements and variation of the calculated available bandwidth. The beats are caused by the variable part of the delay; its role is reduced as the number of measurements.

In this section the necessary number of measurements is calculated using two methods: from experimental data of the RIPE Test Box and by simulation knowing the distribution type for network delay.

Based on data obtained from tt01 and tt143 Boxes we were computed standard deviations (SD) $\sigma_n(B)$ of available bandwidth.

\begin{figure*}
\centering
\includegraphics[height=7.5cm]{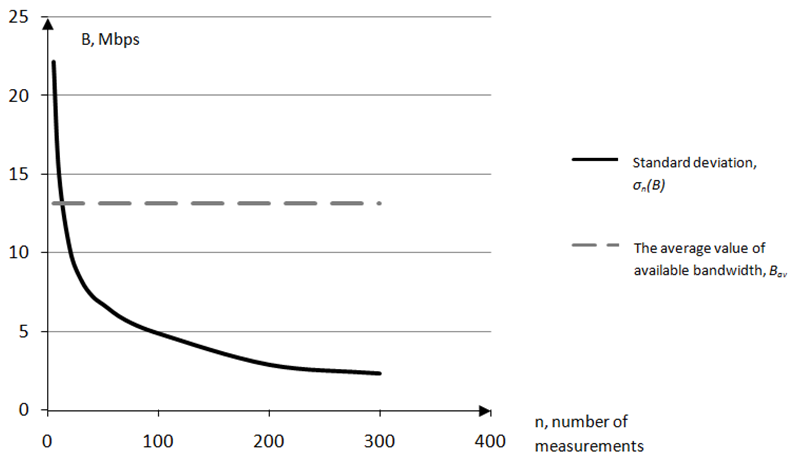}
\caption{Dependence of SD on the number of measurements}
\label{f4}
\end{figure*}

Data are presented in Table~\ref{t4} and graphically depicted in Fig.~\ref{f4}.

\begin{table*}
	\centering
		\begin{tabular}{|c|c|c|c|c|c|c|c|c|c|c|}
		\hline
Number of measurements, $n$	& 5	& 10	& 20	& 30	& 40 & 50 & 70 & 100 & 200 & 300\\
		\hline
Standard deviations, & \multirow{2}*{22.2}	& \multirow{2}*{14.9} &	\multirow{2}*{10.2}	& \multirow{2}*{8.3} &	\multirow{2}*{7.3}	& \multirow{2}*{6.7} &	\multirow{2}*{5.7} &	\multirow{2}*{4.9}	& \multirow{2}*{2.9}	& \multirow{2}*{2.3}
\\ $\sigma_n(B)$ ({\em Mbps}) & & & & & & & & & & \\	
		\hline
The average value & \multicolumn{10}{|c|}{}
\\ of available bandwidth, & \multicolumn{10}{|c|}{13.1}
\\ $B_{av}$ ({\em Mbps}) & \multicolumn{10}{|c|}{}\\
		\hline
		\end{tabular}
	\caption{Dependence of SD on the number of measurements}
	\label{t4}
\end{table*}

Figure 4 shows that it is necessary to take at least 50 measurements (the delay difference for 50 pairs of packets). In this case, the calculated value exceeds twice the capacity of SD, i.e.: $B\geq 2\sigma_n(B)$.

A more accurate result can be obtained using the generating functions for describing the delay packets. In paper \cite{s3} it is shown that the delay distribution is described by exponential law and the following generating function can be used for delay emulation:
 	\begin{equation}
  D=D_{min}+W/B-(1/\lambda)ln(1-F(D,W)),
  \label{e7}
	\end{equation}
where $\lambda=1/(D_{av}-D_{min})$. The function $F(W,D)$ is a standard random number generator in the interval $[0;1)$.

Knowledge of the generating function allows calculating the tabulated values of $\eta^{T}_{n}$ from Eqn.~\ref{e5}. Earlier standard deviation $\sigma^{T}_{n}(D_2-D_1)$ for the delay difference is found taking $\lambda^T=1000 s^{-1}$. Calculation will hold for the following values: $\Delta W^T=W_2-W_1=1000$ {\em bytes}, $B^T=10$ {\em Mbps}, which corresponds to $D^{T}_{2}-D^{T}_{1}=8\cdot 10^{-4}${\em s}.

\begin{table*}
	\centering
		\begin{tabular}{|c|c|c|c|c|c|c|c|}
		\hline
Number of measurements, $n$	& 5	& 10	& 20	& 30	& 50 & 100 & 200\\
		\hline
Standard deviations, & \multirow{2}*{0.661}	& \multirow{2}*{0.489} &	\multirow{2}*{0.354} &	\multirow{2}*{0.284}	& \multirow{2}*{0.195}	& \multirow{2}*{0.111} &	\multirow{2}*{0.075}
\\ $\sigma^{T}_{n}(D_2-D_1)$ ({\em ms})& & & & & & & \\
		\hline
		\end{tabular}
	\caption{Emulation of dependence SD on the number of measurements ($\lambda^T=1000$ {\em s}$^{-1}$)}
	\label{t5}
\end{table*}

For the $\sigma^{T}_{n}(D_2-D_1)$ values from Table~\ref{t5} values of $\eta^{T}_{n}$ could be found (see Table~\ref{t6}).

\begin{table}[!h]
	\centering
		\begin{tabular}{|c|c|c|c|c|c|c|c|}
		\hline
Number $n$	& \multirow{2}*{5}	& \multirow{2}*{10}	& \multirow{2}*{20}	& \multirow{2}*{30}	& \multirow{2}*{50} & \multirow{2}*{100} & \multirow{2}*{200}
\\ of measurements, & & & & & & & \\
		\hline
Measurement &	\multirow{2}*{82.6} &	\multirow{2}*{61.1} &	\multirow{2}*{44.2} &	\multirow{2}*{35.5} &	\multirow{2}*{24.4} &	\multirow{2}*{13.9} &	\multirow{2}*{9.4}
\\ error, $\eta^{T}_{n}$ (\%)& & & & & & & \\
		\hline
		\end{tabular}
	\caption{Dependence of error on the number of measurements}
	\label{t6}
\end{table}

During the real experimental measured quantities $\lambda^{exp}$, $D^{exp}_{2}-D^{exp}_{1}$, and $B^{exp}$ take arbitrary values, but correction factors can easily calculate the required number of measurements:

 	\begin{equation}
  \eta^{T}_{n}=k(D_2-D_1)\cdot k(\lambda)\cdot \eta^{exp}_{n},
  \label{e8}
	\end{equation}
where $k(\lambda)=\lambda^{exp}/\lambda^T$, and $k(D_2-D_1)=(D^{exp}_{2}-D^{exp}_{1})/(D^{T}_{2}-D^{T}_{1})$.

Substituting in Eqn.~\ref{e8} values of the coefficients $k(D_2-D_1)$, $k(\lambda)$ and the desired accuracy of measurements $\eta^{exp}$ we compare the obtained values with the tabulated $\eta^{T}_{n}$ and find the number of measurements $n$ required to achieve a given error.

\section{Conclusion}
In this paper we found a way to measure available bandwidth by data of delays that could be collected by RIPE Test Box. This method consists in the fact that comparing the average end-to-end delays for packets of different sizes, we can calculate the available bandwidth.

We carried out a further study of the model and found the limits of its applicability, which depend on the accuracy of measurement delays. The experiment results were obtained with our method and the alternative one. The benchmark tool has been selected utility \textit{pathload}. The paper shows that the accuracy of calculations available bandwidth depends on variable delay component, $d_{var}$.
 
The experiments and computer simulation, using the generating function of the delay were conducted. They have shown that achieving a given error requires to average large number of measurements. We found a relationship between accuracy and the number of measurements to ensure the required level of accuracy.

In the future we plan to implement this method in mechanism of the measurement infrastructure RIPE Test Box.

\section*{Acknowledgment}

We would like to express special thanks Prasad Calyam, Gregg Trueb from the University of Ohio, Professor Richard Sethmann, Stephan Gitz and Till Schleicher from Hochschule Bremen University of Applied Sciences, Dmitry Sidelnikov from the Institute of Organic Chemistry at the Russian Academy of Sciences for their invaluable assistance in the measurements. We would also like to express special thanks all the staff of technical support RIPE Test Box especially Ruben fon Staveren and Roman Kaliakin for constant help in the event of questions concerning the measurement infrastructure.

\end{document}